\documentclass[12pt,preprint]{aastex}
\usepackage{natbib}
\begin{document}
\title{High-Contrast 3.8 Micron Imaging Of The \\
Brown Dwarf/Planet-Mass Companion to GJ 758}
\author{Thayne Currie\altaffilmark{1}, Vanessa Bailey\altaffilmark{2}, 
Daniel Fabrycky\altaffilmark{3}, Ruth Murray-Clay\altaffilmark{3}, 
Timothy Rodigas\altaffilmark{2}, Phil Hinz\altaffilmark{2}
}
\altaffiltext{1}{NASA-Goddard Space Flight Center}
\altaffiltext{2}{Steward Observatory, University of Arizona}
\altaffiltext{3}{Harvard-Smithsonian Center for Astrophysics}
\begin{abstract}
We present L' band (3.8 $\mu m$) MMT/Clio high-contrast imaging data for the nearby star GJ 758, which was recently 
reported by Thalmann et al. (2009) to have one -- possibly two-- faint comoving companions 
(GJ 758B and ``C", respectively).  GJ 758B is detected in two distinct datasets.
Additionally, we report a \textit{possible} detection of the object identified by 
Thalmann et al as ``GJ 758C" in our more sensitive dataset, though it is likely 
a residual speckle.  However, if it is the same object as that 
reported by Thalmann et al. it cannot be a companion in a bound orbit.  
%Comparing these data to previous astrometric measurements indicates that GJ 758B 
%exhibits counterclockwise motion consistent with Keplerian rotation.
GJ 758B has a H-L' color redder than nearly all known L--T8 dwarfs.
%slightly redder than but 
%comparable to similarly aged Wolf 940B, a 20--30 M$_{J}$ T8.5 dwarf with 
%T$_{e}$ = 570 K.
%If our detection of GJ 758C is indeed real, its inferred 
%H-L' color is more consistent with being closer to the L/T dwarf boundary.  
Based on comparisons with the COND evolutionary models,  
 GJ 758B has T$_{e}$ $\sim$ 560 K$^{^{+150 K}_{-90K}}$ and a mass ranging from $\sim$ 10--20 M$_{J}$ 
if it is $\sim$ 1 Gyr old to $\sim$ 25--40 M$_{J}$ if it is 8.7 Gyr old.
GJ 758B is likely in a highly eccentric orbit, e $\sim$ 0.73$^{^{+0.12}_{-0.21}}$, with a 
semimajor axis of $\sim$ 44 AU$^{^{+32 AU}_{-14 AU}}$. 
Though GJ 758B is sometimes discussed within the context of exoplanet direct 
imaging, its mass is likely greater than the deuterium-burning limit 
and its formation may resemble that of binary stars rather than 
that of jovian-mass planets.
\end{abstract}
\section{Introduction}
High-contrast imaging surveys have recently identified many faint, cool companions 
to nearby stars whose inferred masses are between that of Jupiter and the 
deuterium burning limit ($\sim$ 13 M$_{J}$).
  The companions to nearby A stars -- HR 8799, Fomalhaut, and $\beta$ Pic --
orbit at separations less than $\sim$ 100 AU, have small mass ratios, and generally resemble scaled-up 
versions of gas giant planets in our solar system \citep{Marois2008,Kalas2008,Lagrange2010},  
with temperatures that are (likely) $\lesssim$ 1500 K\citep{Marois2008,Hinz2010,Currie2010}, comparable to that 
for many L and T dwarfs \citep[e.g.][]{Metchev2006,Leggett2010}.
\footnote{Dynamical constraints for the masses of some of these companions -- e.g. those orbiting HR 8799 -- 
are consistent with luminosity-derived estimates \citep{Chiang2009,Fabrycky2010}}.  
Other directly imaged companions have similar temperatures but orbit low-mass stars and brown 
dwarfs, typically at wider separations and/or with much larger mass ratios \citep[e.g.][]{Chauvin2004,Itoh2005,Luhman2006,
Lafreniere2010,Todorov2010}, indicating that they plausibly represent the low-mass tail of 
objects formed by molecular cloud core or protostellar disk fragmentation \citep[e.g.][]{Lodato2005}.
The planetary companions to A stars pose strong challenges for 
even the most efficient core accretion models of gas giant planet formation, given the difficulty of 
 forming massive cores at 10--100 AU \citep[e.g.][]{Rafikov2010} prior to the dispersal of 
the protoplanetary disk.  The configuration of at least one system with a low-mass primary 
and planet-mass companion, 2MASS J04414489+2301513, implies that the fragmentation of 
 molecular cloud cores can produce objects $\lesssim$ 5--10 M$_{J}$, below the 
classical "opacity-limited" minimum fragmentation mass \citep{Todorov2010}.

The faint companion and candidate companion to the nearby star GJ 758 
reported by Thalmann et al. (2009; hereafter T09), GJ 758B and ``GJ 758C", 
 present an intriguing contrast to planets orbiting A stars and 
to other substellar-mass L/T dwarfs.  GJ 758 is much 
later in spectral type (G8V) and lower in mass (M$_{\star}$ = 0.97 $\pm$ 0.03 
M$_{\odot}$, \citealt{Takeda2007}) than HR 8799, Fomalhaut, and $\beta$ Pic 
(1.5--2.1 M$_{\odot}$).  On the other hand, the projected separation 
for GJ 758B (1.858", PA = 198.18$^{o}$ on 2009 Aug. 6, or $\sim$ 29 AU) 
and ``GJ 758C" (1.118", PA = 219.16$^{o}$, or $\sim$ 18 AU) are comparable to the  
separations for planets orbiting HR 8799 and $\beta$ Pic 
\citep{Marois2008,Lagrange2010} and within the brown dwarf desert \citep[e.g.][]{Kraus2008}.
Based on GJ 758B's H band magnitude and age (0.7--8.7 Gyr), 
T09 argue that the companion likely has a temperature of 549--637 K 
(M = 11.7--48.5 M$_{J}$), making it the coldest known 
companion to a Sun-like star.  Since T09 image 
the system only in H band, the companion's temperature  --and thus 
luminosity -- remain observationally unconstrained.  
  Inferred masses critically depend on these properties, so new longer wavelength 
data help constrain whether GJ 758B better resembles the low mass-ratio 
 planetary companions to A stars or high mass-ratio 
binary star companions to lower-mass stars.
 
Here we report the L' band detection of GJ 758B and candidate detection 
of ``GJ 758C" with the Clio camera at the 6.5m MMT telescope.  Combined 
with the H band photometry from T09, we 
estimate the temperature, luminosity, and mass of GJ 758B.   Combining 
all astrometric datapoints for GJ 758B, we determine its range of allowed orbits.

\section{Observations and Data Reduction}
GJ 758 was imaged under photometric conditions on May 27, 2010 and June 2, 2010 at the 
6.5m MMT telescope with the upgraded Clio mid-IR camera \citep{Hinz2006,Sivanandam2006}.
As described in online documentation\footnote{http://zero.as.arizona.edu/wiki/doku.php}, 
the new Clio detector (Clio-2) operates from 1.65 $\mu m$ to 4.8 $\mu m$, covering standard 
Mauna Kea filters and narrowband filters centered on wavelengths between 3 and 4 $\mu m$.  
Both sets of data reported here were obtained in the L' filter (3.8 $\mu m$).
All data was taken in \textit{angular differential imaging} mode \citep{Marois2006}, which 
keeps the instrument rotator fixed, allowing the field of view to rotate with time. 
The May data consist of coadded frames of 9s each for a total 
integration time of 2520s.  The data were taken through transit, yielding 
a total field rotation of 192 degrees.  The June data consist of coadded 
frames of 9.6s each for a total integration time of 1536s 
but were obtained well after transit for a total field rotation of 16.7 degrees.  
In both datasets, the star 
was dithered along the detector by 5--10" every 48--90s for sky subtraction.  
For precise astrometric calibration during our run, we observed the double stars 
HIP 88817/88818 and HD 223718.
The detector orientation is offset by 2.53$^{o}$ $\pm$ 0.15$^{o}$ counterclockwise from true north along the 
y axis; the pixel scale is 0.029915 $\pm$ 8$\times$10$^{-5}$ "/pixel.   
Independent astrometry performed by multiple coauthors confirm these values.
Based on this pixel scale, our image FWHM is $\sim$ 0.15" for both datasets.

Our image processing method closely followed the ADI/LOCI (Angular Differential 
Imaging/Locally-Optimized Combination of Images) reduction procedure described 
by \citet{Marois2006,Marois2008} and \citet{Lafreniere2007} used  
to maximize the companion signal relative to speckle-dominated noise.
The subtracted residual images produced by our ADI/LOCI pipeline were then derotated, median 
combined, and convolved with a gaussian kernel  equal to the image FWHM to produce 
final science images.  Separately from our LOCI reduction, we performed 
a simple ADI reduction by derotating, high-pass filtering, and median 
combining the PSF subtracted images \citep[e.g.][]{Hinz2010}. 

\begin{figure}
%\centering
\includegraphics[trim = 0mm 0mm 7mm 7mm,clip,scale=0.8]{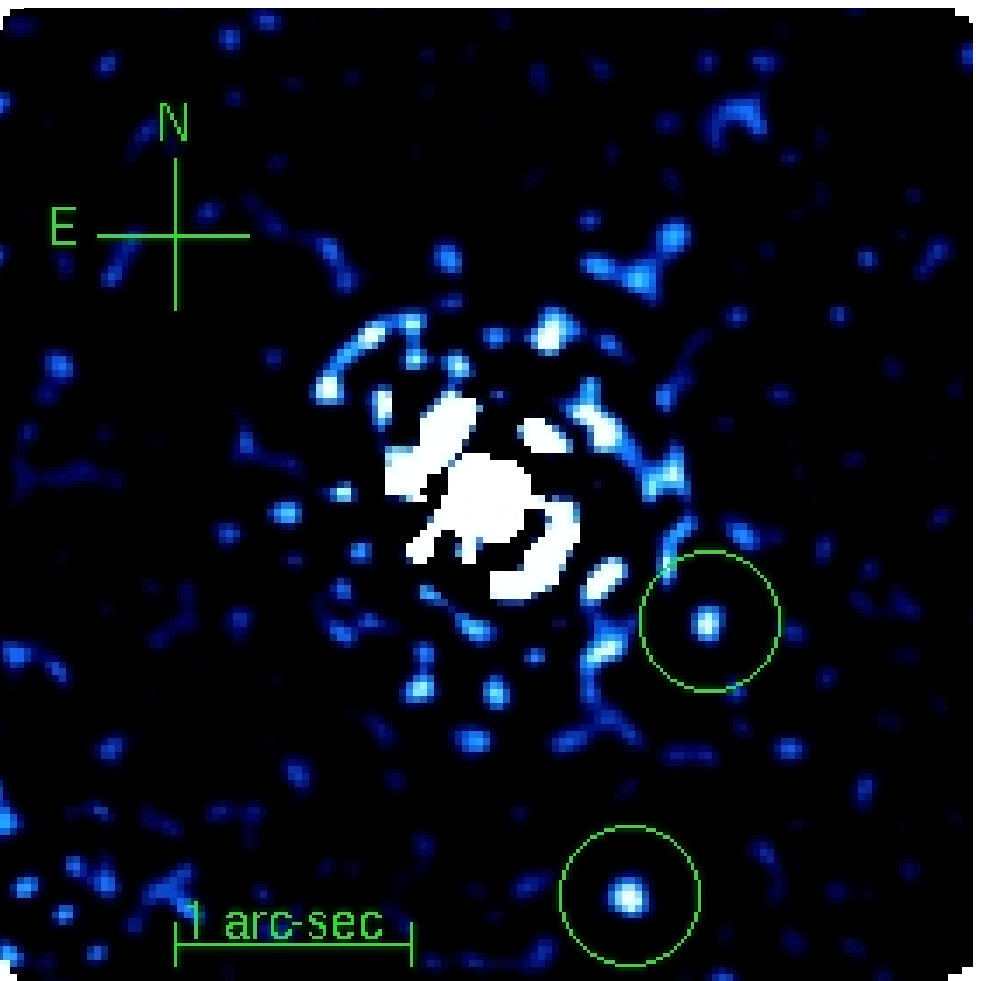}
\includegraphics[trim = 0mm 0mm 7mm 7mm,clip,scale=0.8]{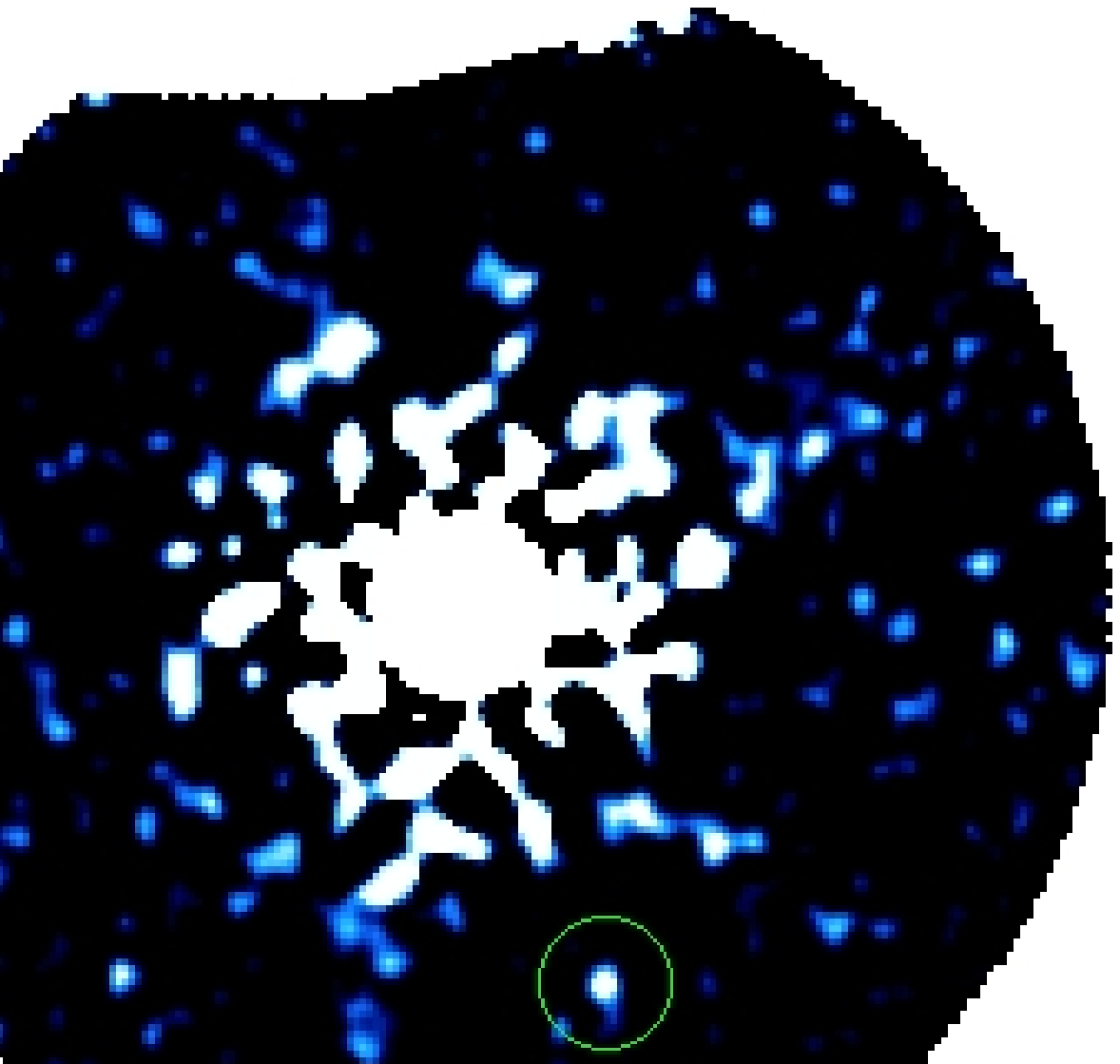}
\caption{LOCI-reduced images from our May dataset (left) and 
 June dataset (right) shown at high contrast to display residual speckle noise.
  GJ 758B (lower circle) is recovered in both 
data.  A candidate point source with a separation comparable to that reported for 
``GJ 758C" (top circle, left panel) is also detected in the more sensitive May data, 
though this may instead be a residual speckle.  Because of shorter 
integration times and far poorer field rotation, residual speckle noise 
is far more severe for the June data, and we do not recover the second, candidate point source.}
\label{images}
\end{figure}

Figure \ref{images} shows the final reduced images for the 
May dataset (left) and June dataset (right) produced with the 
LOCI pipeline.  The May dataset clearly reveals a point source at a separation 
of 1.823" $\pm$ 0.015" (a$_{proj}$ $\sim$ 28.4 AU) and position angle of 199.76 $\pm$ 0.15 degrees.
Within astrometric errors, we identify a point source at
the same position in the lower-quaility June data (1.827" $\pm$ 0.043"\footnote{The much larger 
error bars result because the companion is contaminated by some residual speckle noise}), 
indicating that its detection is secure.  We recover the point source in our more simple ADI 
reductions as well.  This separation is comparable to that for 
GJ 758B as reported by T09 (1.858" $\pm$ 0.005"); the difference in 
position angle ($\delta$(PA) = 1.58$^{o}$) is consistent with an object undergoing 
counterclockwise motion with respect to the primary.  
The velocity in the plane of the sky of GJ 758B, 
derived by comparing our May 2010 position 
with previously reported positions (T09), is 
1.3 $\pm$ 0.2 AU/yr. This is less than the minimum escape velocity 
at this projected separation (1.64 $\pm$ 0.03 
AU/yr$\times$(M$_{\star}$/0.97M$_{\odot}$)$^{1/2}$$\times$(a$_{p}$/28.50 AU)$^{-1/2}$), 
suggesting a bound orbit.  Thus, we conclude that we have detected GJ 758B.

The LOCI reduction of the May data also reveals a 
second, candidate point source at 1.101" $\pm$ 0.015", 
 comparable to that reported for 
the candidate companion, ``GJ 758C", in T09.
However, the large difference in position angle ($\sim$ 20$^{o}$) 
would imply a space velocity of 7.9 $\pm$ 0.3 AU/yr, nearly four times greater than 
the escape velocity of 2.08 AU/yr.
  Thus, if our candidate point source is ``GJ 758C", 
it cannot be in a bound orbit around the primary.

To assess the significance of the GJ 758B detection and candidate ``GJ 758C" 
detection, we compute the standard 
deviation and signal-to-noise of pixel values in concentric annuli (e.g. T09).  
The signal-to-noise is $\sim$ 6.4 
at GJ 758B's position and $\sim$ 3 at the candidate ``GJ 758C" position.
Thus, the signal-to-noise of our possible ``GJ 758C" detection is only marginally significant 
($\sim$ 3$\sigma$), even though it is locally well separated from large background 
fluctuations.  Bona-fide detections in speckle-noise limited regions typically need to be 
greater than 5$\sigma$ in order to rule out false "detections" from residual speckles 
\citep{Marois2007}. 
Because of the chance that ``GJ 758C" could be a residual speckle and the lack of a detection in our June data, 
we consider the second point source in our May data to only be a \textit{candidate} detection of the 
 point source identified by T09 as ``GJ 758C" \footnote{Chance superpositions of substellar objects close to 
stars like GJ 758 with L' band brightnesses similar to our candidate detection 
are unlikely to be frequent.  
%Assuming a number density of $\approx$ 6000 point sources deg$^{-2}$ 
%at 3--4 $\mu m$ brighter than magnitude 16.5 \citep{Fazio2004}, we are unlikely to 
%encounter a detectable background object within 2" of the primary (Prob(bckgd) $\sim$ $\pi$2$^{2}$$\times$3000/(3600$^{2}$) 
%= 0.58\%).  
Assuming a flat 
substellar mass function ($\alpha$ = 0) with a space density described in Table 4 of \citet{Burgasser2004} 
using the absolute magnitude vs. spectral type of L/T dwarfs at the Spitzer/IRAC [3.6] band 
from \citet{Leggett2010} as a proxy for L' brightness, the probability 
of contamination by a background brown dwarf with T$_{e}$ =500--1500 K is $<$ 10$^{-3}$ $\%$.  
%source counts are also low at GJ 758's galactic longitude (66.0765 degrees) \citep{Churchwell2009}.}.  
}
  Since only the detection of GJ 758B is secure, we focus on it in our analysis.

\section{Photometric and Astrometric Analysis}
%\subsection{Photometry and Temperatures of GJ 758B}
Photometry for GJ 758B from the May data was performed with IDLPHOT, using a 2.5 pixel aperture radius 
and a background annulus between 2.5 and 5 pixels.  In all exposures, the stellar PSF core 
is saturated.  For photometric calibration, we compare the GJ 758B flux to that for 
HD 223718A, which was observed immediately after GJ 758.
HD 223718A is an F5V star (Te $\sim$ 6440 K, \citealt{Currie2010b}) and should have 
K-L' = 0.04 \citep{Tokunaga2000}.  Based on its 2MASS photometry (K$_{s}$= 6.51) 
and using the updated version of the 2MASS color transformations between 
2MASS and CIT systems \citep{Carpenter2001}, its estimated L' magnitude is 6.49. 
Both the ADI and LOCI reduction procedures attenuate some 
companion flux in the process of attenuating speckles.  
To further calibrate our photometry, we introduce and measure 
the flux for fake point sources at random angles at a range of separations (0.5"--2.5") in 
each registered frame, rerun our ADI and LOCI pipelines, compute the attenuated
flux in the final ADI/LOCI-processed images, and correct for this attenuation 
($\sim$ 30\% at GJ 758B's position for the LOCI reduction, $\sim$ 10\% for ADI).  
 GJ 758B has L' = 15.97 $\pm$ 0.19, where 
our photometric errors accounts for 1) the intrinsic photometric 
uncertainty for GJ 758B, 2) the intrinsic photometric uncertainty 
for our calibration star, and 3) the differences in photometry 
between the ADI and LOCI-reduced images\footnote{If if our 
second, candidate point source is indeed the object 
T09 identify as ``GJ 758C", its magnitude would be L' = 16.01 $\pm$ 0.38; the 3$\sigma$ 
limit at its location reported in T09 is 
$\sim$ 16.03.  Regardless of whether our candidate 
detection of ``GJ 758C" is real or is a speckle, the H-L' color for the purported 
``GJ 758C" object must be bluer than $\approx$ 2.5.}.  
Using the T09 H band photometry, we then find H-L'= 3.29 
$\pm$ 0.25 for GJ 758B.

The H-L' color provides a first-order estimate for GJ 758B's temperature 
and spectral type (Table 1).  GJ 758B is far redder than than expected 
if it were a L--T8 dwarf\footnote{See the sample compiled by S. Leggett: \url[HREF]{http://staff.gemini.edu/$\sim$sleggett/LTdata.html}}.  
However, GJ 758B is comparable in color to Wolf 940B (H-L' = 3.38),  
 a 3.5--6 Gyr-old, 570 K T8.5 dwarf companion to a higher-mass brown dwarf \citep{Burningham2008}.  
To derive GJ 758B's effective temperature, we compare its colors to the colors predicted by the 
COND evolutionary models for its age range \citep{Baraffe2003}.  Based on isochronal fitting and 
age-activity correlations \citep[e.g.][]{Takeda2007, Mamajek2008}, GJ 758 has an 
age between 0.7 Gyr ($\sim$ 1 Gyr) and 8.7 Gyr (see discussion in T09).
The 0.7 Gyr lower limit is based on isochrone fitting from the Y$^{2}$ tracks 
\citep{Takeda2007}, which may be highly uncertain given that GJ 758 
is on/near the main sequence and given systematic disagreements between different 
sets of isochrones.  For simplicity, we assume an age range of 1 Gyr to 10 Gyr to 
cover the low end of the age range and the main sequence lifetime of 
a Sun-like star \citep{Sackmann1993}.
For this age range, GJ 758B has an allowed temperature range of 
log(T$_{e}$) $\sim$ 2.675-2.85 or T$_{e}$ = 560$^{^{+150}_{-90}}$ K, where the extrema 
are determined from the $\pm$ 1$\sigma$ values for the H-L' color (3.04 and 3.54)  
and a slight age dependent calibration between temperature and color. 

The implied luminosity of GJ 758B as probed by the companion's absolute H band magnitude 
and H-L' color is broadly consistent with COND model predictions of GJ 758B's age range.  
For GJ 758B's H-L' color, the COND models at 5 and 10 Gyr predict an absolute H band 
magnitude of $\approx$ 18.4 and 18.2, respectively.  At 1 Gyr, the COND models 
predict M$_{H}$ $\approx$ 18.6, though given the uncertainty in the H-L' color
the predicted near-IR absolute magnitudes are still consistent with observations.  
Given our range in effective temperatures and comparing L/L$_{\odot}$ and Te from 
1--10 Gyr for the COND models, the bolometric luminosity of 
GJ 758B is log(L/L$_{\odot}$) $\approx$ -6.1$^{^{+0.3}_{-0.2}}$. 

\begin{figure}
\plotone{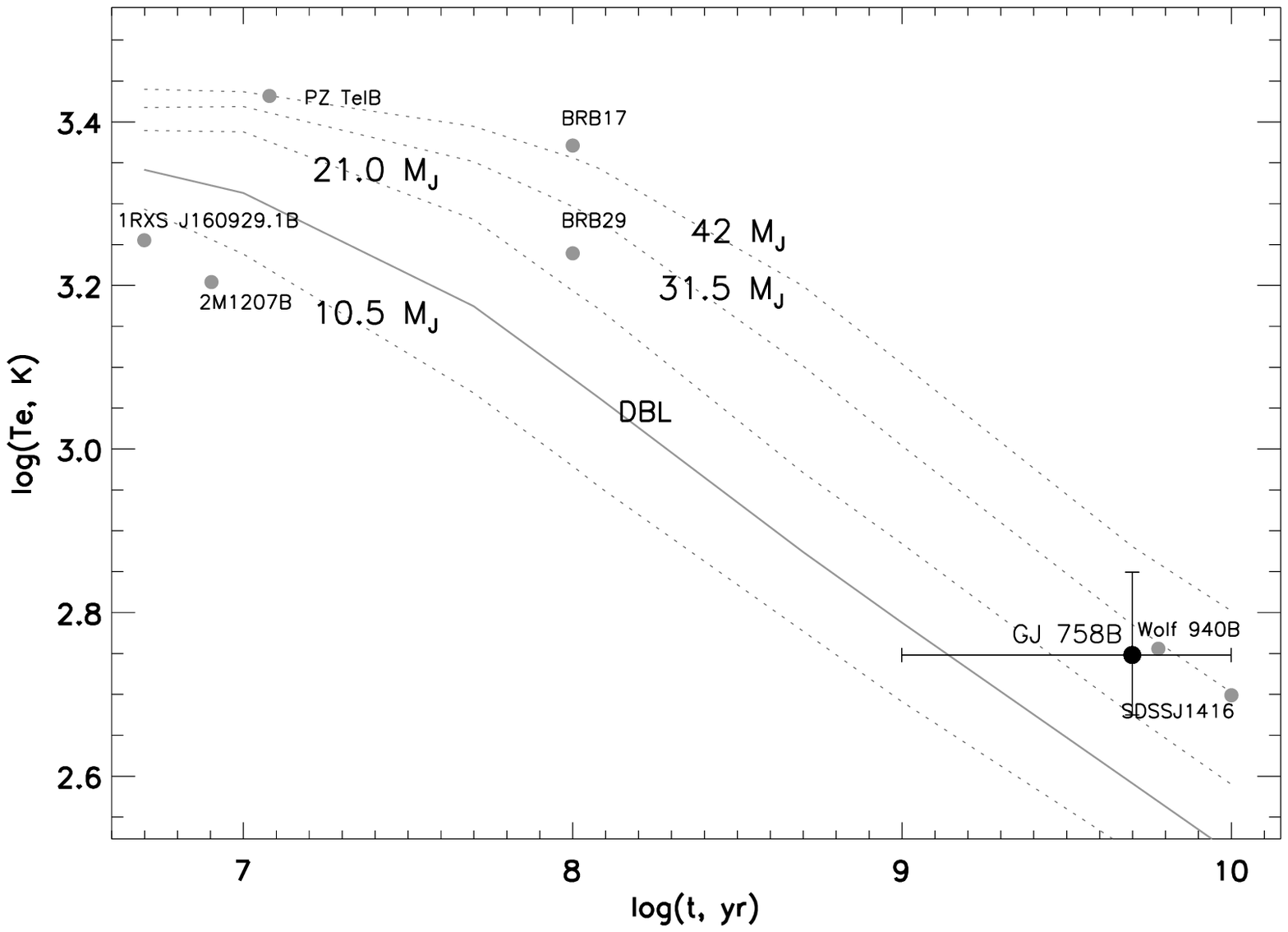}
\caption{Effective temperature vs. age for low-mass brown dwarfs/planetary-mass objects orbiting 
solar and subsolar-mass stars.  Along with GJ 758B, these include PZ Tel \citep{Biller2010}, 1RXS J160929.1B \citep{Lafreniere2008,Lafreniere2010}, 
2M 1207B \citep{Chauvin2004}, two L dwarfs in the Pleiades \citep{Bihain2010}, Wolf 940B \citep{Burningham2008}, 
and SDSS J1416+13AB \citep{Burningham2009}.  We overplot the temperature evolution 
for 10.5--42 M$_{J}$ objects from the COND evolutionary models \citep{Baraffe2003}. 
}
\label{tempevo}
\end{figure}

Figure \ref{tempevo} plots the predicted temperature evolution of 10.5--42 M$_{J}$ 
companions compared to the temperatures of GJ 758B and other cool, substellar-mass 
objects above and below the deuterium-burning limit.  
GJ 758B has an inferred mass ranging from 10--20 M$_{J}$ (at 1 Gyr) to $\sim$ 
25--40 M$_{J}$ (at 10 Gyr), consistent with estimated mass for Wolf 940B (24--45 M$_{J}$). 
Within the context of the temperature evolution of substellar objects 
predicted by the COND models, GJ 758B is likely higher mass than 1RXS J162041-210524B 
\citep{Lafreniere2008,Lafreniere2010} and 2M 1207B \citep{Chauvin2004}.
GJ 758B is consistent with being an older, cooler analogue to mid-L brown dwarfs 
in the Pleiades (e.g. BRB29) studied by \citet[][and references therein]{Bihain2010}  
and the recently discovered brown dwarf companion to PZ Tel, a 12 Myr-old 
1.3 M$_{\odot}$ star \citep{Biller2010}.  

For most of the GJ 758B age range, the COND models yield estimated masses 
above the deuterium burning limit nominally separating planets from brown dwarfs.
The more reliable age indicators, age-activity relations 
\citep[e.g.][]{Mamajek2008,Barnes2007}, suggest that GJ 758 at least as old as the 
Sun, and perhaps as old as $\sim$ 8.7 Gyr (T09).  Thus, the mass of GJ 758B is most plausibly comparable 
to that of low-mass brown dwarfs, not high-mass planets.

\begin{figure}
\plottwo{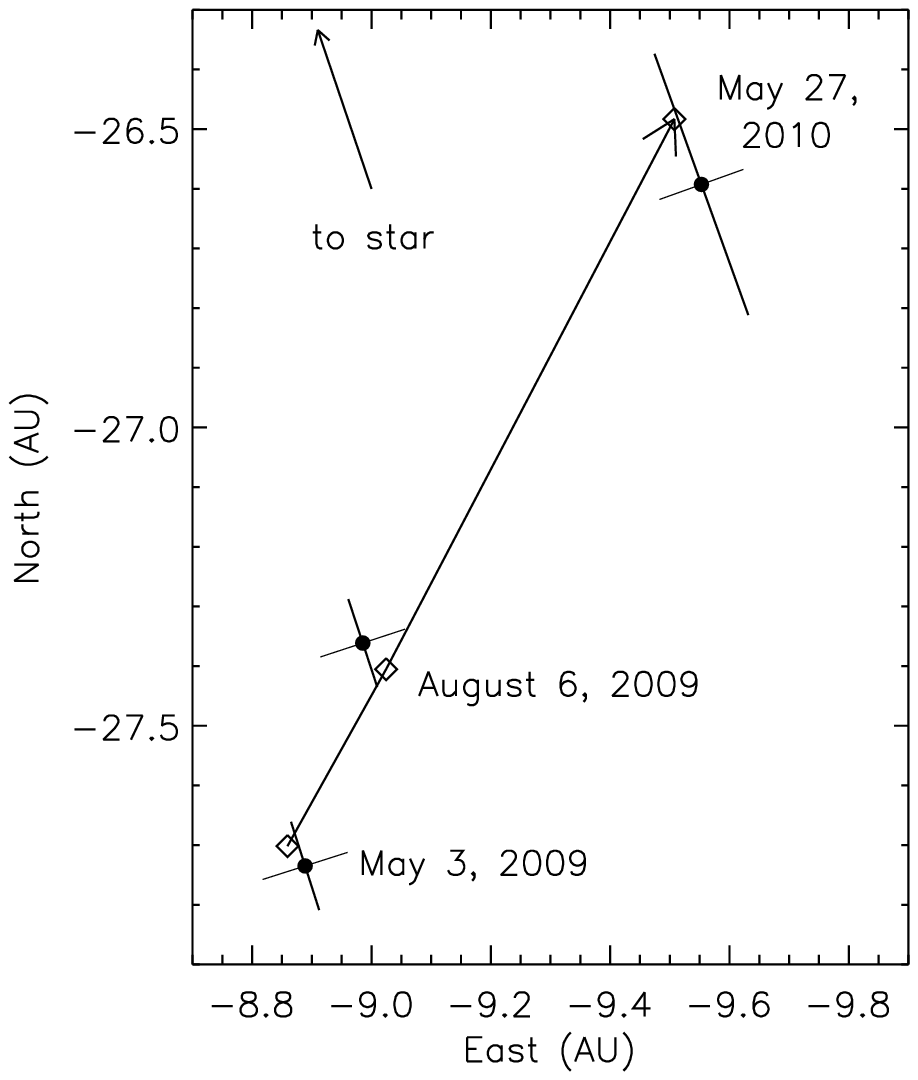}{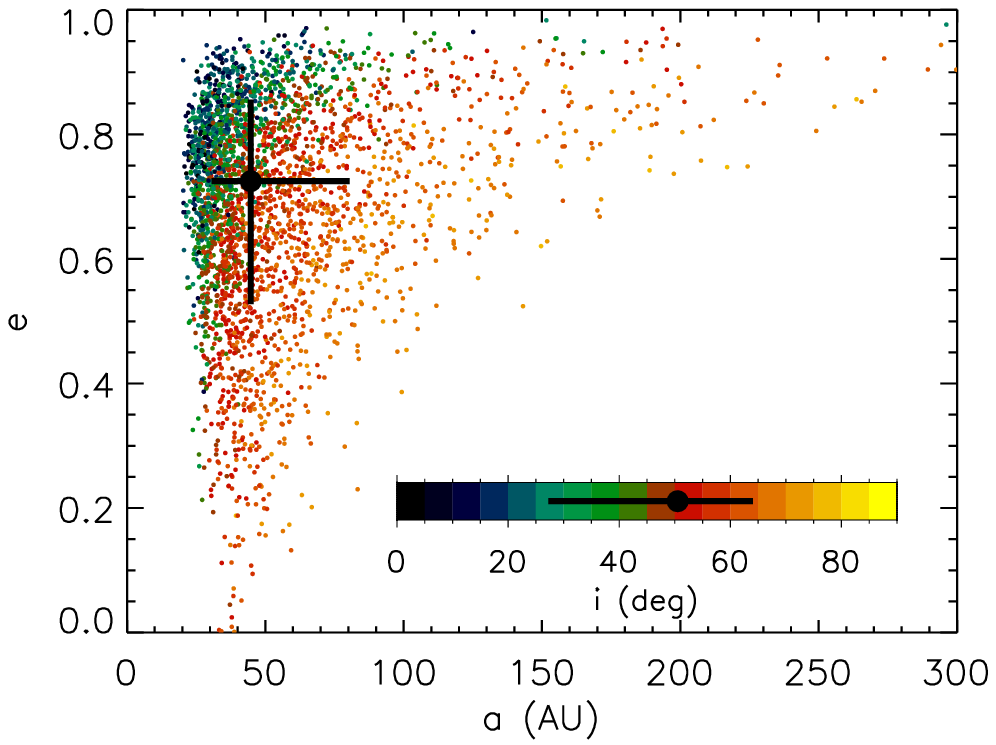}
\caption{(left) Projected separation for GJ 758B between May 2009 and May 2010, 
combining our data (upper right) with that from T09.  The arrows show the motion of GJ 758B from May 2009 (lower left) 
to May/June 2010 (upper right); the diamonds are the best-fit positions based on a rectilinear model.  (right) 
Range of orbital solutions for GJ 758B following the method of T09.}
\label{astrom}
\end{figure}

We investigate the range of plausible orbital parameters with a Monte Carlo simulation
following the method of T09.   With three astrometric points spanning 
one year (Figure \ref{astrom}, left panel), the orbital properties of GJ 758B are 
better constrained: only 1.2\% of the trial orbital solutions "fit" the data 
compared to $\sim$ 6\% based on only the 2009 data (T09). 
The best-estimated eccentricity is high (e$_{M.W.}$ $\sim$ 0.73$^{^{+0.12}_{-0.21}}$), and the best-estimated 
inclination indicates that the orbit is viewed neither pole on nor edge on(i$_{M.W.}$ $\sim$ 
50$^{o}$$^{^{+14^{o}}_{-22^{o}}}$).  The best-estimated semimajor axis is $\sim$ 44 AU$^{^{+32 AU}_{-14 AU}}$, or 
$\approx$ 60\% larger than its projected separation of $\sim$ 28 AU.  

\section{Discussion}
Our study recovers the detection of GJ 758B reported by T09 and 
demonstrates that it has a red H-L' color consistent with being an ultra-cool (T $\approx$ 560 K)
low-luminosity companion to a solar-type star.  This estimate is consistent with the 
low end of the temperature range (550--640 K) quoted by T09 based on assuming that 
the COND models accurately reproduce the fluxes of substellar mass objects in 
each near-IR bandpass, although the allowed temperature range is about the same 
as that quoted by T09, given our photometric errors.  We also identify a second point source at a separation 
comparable to the putative ``GJ 758C" companion reported by T09, though 
its detection has a low significance and it is likely a residual speckle. 
However, if this is the same object as that previously reported, 
it is not in a bound orbit around GJ 758.

Based on the COND evolutionary models, GJ 758B has a mass between $\sim$ 10 M$_{J}$ and 
40 M$_{J}$.  However,  for most of GJ 758's age range, especially for ages derived from 
more reliable methods, its inferred mass is greater than the deuterium burning limit.    
Thus, based on our analysis, it is then much more likely that GJ 758B is a low-mass brown dwarf companion 
than a planet if "planet" is defined by an object's mass relative to the deuterium burning limit.  

A separate, perhaps more physically motivated criterion for defining an object to be a "planet" is 
its formation history: whether or not it formed in a protoplanetary disk around a young star.
Sophisticated numerical models for planet formation \citep[][]{KenyonBromley2009} 
necessary for testing formation theories 
have yet to be directly applied to systems with directly-imaged planets. 
However, analytical arguments show that forming GJ 758B by core accretion is 
possible only if gas accretion is exceptionally efficient. 
Using Equation 14 from \citet{Rafikov2010}, the minimum surface density of solids required 
to trigger core instability within the lifetime of the disk is $\approx$ 6$\times$10$^{-2}$ 
g cm$^{-2}$ for a (long) protoplanetary disk lifetime of 5 Myr \citep{Currie2009}.  
This surface density is comparable to that for \citeauthor{Rafikov2010}'s Minimum Mass Solar 
Nebula profile at our best fit semimajor axis of 44 AU.
Once a core has formed, the isolation mass to which a non-migrating gas giant could grow 
is $>$ 10--20 M$_{J}$ in disks containing that quantity of gas. 
  However, accreting $>$ 10--20 M$_{J}$ of gas in 
a disk that cannot be more than $\sim$ 100 M$_{J}$ in 
mass total (else the disk would be gravitationally unstable) requires the planet 
to accrete very efficiently.   
Companion formation by gravitational instability of the disk, in  
contrast, can only occur while the disk is still accumulating gas from  
the protostellar core, and preferentially forms companions with masses  
more similar to field brown dwarfs than jovian-mass planets \citep[e.g.][]{Kratter2010,Rafikov2005}.

Formation may also be possible earlier, during the fragmentation of the molecular cloud core that 
formed GJ 758: GJ 758B would then comprise the low-mass end of the binary mass function.  
Core fragmentation can probably produce objects much less massive than GJ 758B 
\citep[e.g. 5--10 M$_{J}$][]{Todorov2010}, so we consider this scenario 
to be a plausible one for forming GJ 758B in addition to fragmentation during the protostar phase.  
Future multiwavelength studies of GJ 758B will better constrain its temperature and atmospheric 
properties.  Combined with improved age estimates derived from stellar rotation and 
activity, these data will better place GJ 758B within the context of directly imaged 
 low-mass brown dwarfs useful for investigating the mass function of objects 
approaching the deuterium burning limit.

\acknowledgements We thank the anonymous referee and Adam Kraus for suggestions that strengthened this paper and 
Adam Burgasser, Marc Kuchner, Scott Kenyon, and Jonathan Irwin for other useful discussions.
  T.C. is supported by a NASA Postdoctoral Fellowship, D. F. is supported by a Michelson Fellowship, and 
R. M.-C. is supported by an Institute for Theory and Computation Fellowship.
{}

\begin{deluxetable}{llllllllllllll}
%\rotate
%\documentstyle[10pt]
%SPMquot"(0pt)
%\setlength{\tabcolsep}{0.003in}
%\linewidth{0.1 in}
%\tabletypesize{\tiny}
%\tabletypesize{\scriptsize}
%\tabletypesize{\small}
\tablecolumns{2}
\tablecaption{Observed and Derived Properties for GJ 758B}
\tiny
\tablehead{{Properties}&{GJ 758B}}
\startdata
%Photometry (May 27, 2010)\\
\textbf{Photometry}\\
App. L$'$ & 15.97 $\pm$ 0.19 mag\\
Abs. L$'$ & 14.84 $\pm$ 0.21 mag\\
H-L'    & 3.29 $\pm$ 0.25 mag\\
\textbf{Derived Physical Properties} &\\
Approx. Spectral Type & T8.5--T9 \\
Te (inferred) & $\approx$ 560 +150, -90 K\\
log(L/L$_{\odot}$) (inferred) & -6.1 +0.3,-0.2 \\
Inferred Mass (M$_{J}$) (1 Gyr) & 10--20  & \\
     (5 Gyr)  & 21.0--31.5 & \\
     (10 Gyr)  & 25--40  & \\
\textbf{Astrometry}\\
\textit{May 3, 2009 (T09)}\\
Separation (") & 1.879 $\pm$ 0.005\\
Position Angle($^{o}$) & 197.17 $\pm$ 0.15\\
\textit{August 6, 2009 (T09)}\\
Separation (") & 1.858 $\pm$ 0.005\\
Position Angle($^{o}$) & 198.18 $\pm$ 0.15\\
 \textit{May 27, 2010 (this work)}\\
Separation (") &1.823 $\pm$ 0.015 \\
Position Angle($^{o}$) &199.76 $\pm$ 0.15\\
Projected Separation (AU) & 28.36\\
$\delta$(Sep., ") & 0.035\\ 
$\delta$(PA, $^{o}$)   & 1.58\\
v$_{R}$ & -0.9 $\pm$ 0.2 AU/yr
\\
v$_{PA}$ & 0.94 $\pm$ 0.1 AU/yr
\\
\textbf{Derived Orbital Properties}\\
Semimajor Axis (AU) & 44.12 (30.28, 76.66) \\
Eccentricity, Inclination & 0.73 (0.52, 0.85), 49.64$^{o}$ (27.87$^{o}$,63.34$^{o}$)\\
\enddata
\tablecomments{The changes in separation ($\delta$Sep.) and position angle ($\delta$PA) 
are given between our May 27, 2010 data and the August 6, 2009 data presented by T09.  
The semimajor axis, eccentricity, and inclination listed refers to the median-weighted value and the 
two values defining the inner 68\% of weighted values.  The instantaneous radial and angular 
velocities (v$_{R}$, v$_{PA}$) were determined at an epoch of 31 Dec 2009 (the error-weighted 
mean of our May datum and T09's data) and an assumed distance of 15.5 pc.}
\end{deluxetable}
\end{document}